\documentclass[12pt,preprint]{aastex}   
\newcommand{\eqwhalpha}{EW(H$\alpha$)}
\newcommand{\eqwhbeta}{EW(H$\beta$)}
\newcommand{\halpha}{H$\alpha$}
\newcommand{\hbeta}{H$\beta$}
\newcommand{\hab}{H$\alpha$/H$\beta$}
\newcommand{\msun}{M$_{\odot}$} 
\newcommand{\HI}{H~{\sc i}}
\newcommand{\HII}{H~{\sc ii}}
\newcommand{\te}{T$_e$}
\begin{document}    

\title{Dust In I\,Zw\,18 From Hubble Space Telescope\footnote{Based on 
observations with the NASA/ESA Hubble Space Telescope, obtained at the 
Space Telescope Science Institute, which is operated by the Association 
of Universities for Research in Astronomy, Inc. under NASA contract No. 
NAS5-26555.}\\ Narrow Band Imaging}

\author{John M. Cannon and Evan D. Skillman}
\affil{Department of Astronomy, University of Minnesota, 116 Church St. 
S.E., Minneapolis, MN 55455}
\email{cannon@astro.umn.edu, skillman@astro.umn.edu}
\author{Donald R. Garnett}
\affil{Steward Observatory, University of Arizona, 933 North Cherry Avenue, 
Tucson, AZ 85721}
\email{dgarnett@as.arizona.edu}
\author{Reginald J. Dufour}
\affil{Department of Physics and Astronomy, Rice University, 6100 Main Street, 
Houston, TX 77005}
\email{rjd@rice.edu}

\begin{abstract}

We present new WFPC2 narrow band imaging of the blue compact dwarf galaxy 
I\,Zw\,18, which is host to the lowest-metallicity \HII\ regions known.  
Images at \halpha\ and \hbeta\ are combined with archival broad band 
images to allow the study of the ionized gas distribution and morphology.  
Analysis of the \hab\ flux ratio reveals significant enhancements in some 
areas of both the ``Northwest'' and ``Southeast'' regions of the galaxy, 
with ratios elevated to levels as high as 3.4.  The \hab\ ratio varies 
considerably with position throughout the galaxy.  Comparing this distribution 
with the stellar distribution indicates that the regions of enhanced \hab\ 
ratio are not due to the effects of either collisional excitation or 
underlying stellar absorption, and therefore are most likely interpreted as 
the presence of dust.  This dust has an estimated mass of (2-5)$\times$10$^3$ 
\msun, which is consistent with the IRAS far-IR non-detection.  Under the 
assumption that dust traces the presence of molecular gas, these results 
suggest that the molecular component of the ISM of I\,Zw\,18, which is needed 
to fuel its active star formation, is also very clumpy.  Such a distribution 
would be consistent with the recent FUSE non-detections of diffuse H$_2$.

\end{abstract}						


\section{Introduction}
\label{S1}

Since its discovery \citep{zwi66}, the blue compact dwarf (BCD) galaxy 
I\,Zw\,18 has been an exemplary low metallicity star forming galaxy and 
has undergone remarkable observational scrutiny.  The galaxy hosts the 
lowest metallicity \HII\ regions known in the local universe, with oxygen 
abundances (O/H) $\sim$ 0.02 (O/H)$_{\odot}$ \citep[][hereafter SK93]{ski93}.  
As such, the study of I\,Zw\,18 is essential in various astrophysical arenas, 
including dwarf galaxy evolution, primordial helium abundance measurements, 
and star formation mechanisms.    

With an \HI\ heliocentric radial velocity of 749 km sec$^{-1}$, a distance of 
$\sim$ 10 Mpc is usually assumed for I\,Zw\,18 \citep[e.g.,][]{van98}.
\citet{ost00} suggests a slightly larger value of $\sim$ 12.6 Mpc to account 
for Virgocentric infall.  The precise distance determination of I\,Zw\,18 is 
beyond the scope of this paper; we follow \citet{ost00} in adopting a distance 
of 12.6 Mpc, but note that this condition is uncertain and is worthy of 
extensive future study. 

The low metallicity of the I\,Zw\,18 system has led to the suggestion that it 
is undergoing its first burst of star formation \citep{kun86}.  However, the 
uniformity of the nebular abundances derived by SK93\nocite{ski93} in the two 
main components of the system argues against this hypothesis.  The 
evolutionary status of I\,Zw\,18 is of paramount importance, since the system 
offers the closest known reproduction of pristine conditions and therefore the 
best low metallicity environment in which to study big bang nucleosynthesis at 
low redshift.  Additional physical mechanisms have also been studied in detail 
in I\,Zw\,18 to discern their behavior at very low metallicities \citep[e.g., 
heating and cooling balance,][]{sta99}.

The nature of previous generations of star formation within I\,Zw\,18 has not 
yet been fully characterized.  Interestingly, the nebular C/O ratio was found 
to be elevated in I\,Zw\,18 with respect to other comparably metal-poor 
irregular galaxies and above the predictions of massive star nucleosynthesis 
\citep[][but see also Izotov \& Thuan 1999\nocite{izo99}]{gar97}.  The 
implication of such an observation is that I\,Zw\,18 may have undergone a 
previous episode of star formation; this conclusion is supported by the 
discovery of a population of 0.5-5 Gyr old Asymptotic Giant Branch (AGB) stars 
from HST/NICMOS imaging \citep{ost00}. CMD analysis by Aloisi, Tosi, \& Greggio 
(1999)\nocite{alo99} also suggests the presence of evolved intermediate mass 
stars in the blue loop phase which may be as low as 3-4 \msun, implying ages 
of $\sim$ 0.3-0.6 Gyr.  Photometric investigations thus suggest that I\,Zw\,18 
has an (at least) intermediate age stellar population and is thus not 
evolutionarily young. 

I\,Zw\,18 has been the target of deep HST studies since the launch of the
spacecraft, and numerous archival studies have also been executed.  The present 
new images complement and, where possible, improve on, previous works.  Imaging 
studies of the nebular emission (\halpha, He~{\sc ii}) were carried out by 
\citet[][hereafter HT95]{hun95} and \citet{dem98}.  The former analyzed the 
stellar distribution and derived properties of the ionized gas distribution
(\HII\ region characteristics, age arguments).  The latter used He~{\sc ii} 
emission to characterize the population of Wolf-Rayet stars within the galaxy.  
This study adds to these works, in that we have deeper, higher resolution 
\halpha\ imaging, and combine this with \hbeta\ imaging to derive reddening 
characteristics within the galaxy.  This is the first imaging diagnostic of the 
internal extinction using HST data.  Previous evidence for dust in I\,Zw\,18 
has come from several ground-based spectroscopic investigations wherein 
reddening is detected \citep[SK93;\nocite{ski93}][]{izo98,vil98,leg00}.  The 
high spatial resolution afforded by these WFPC2 images allows us to explore the 
nature and spatial distribution of the dust and to address the possibility that 
the elevated \hab\ ratios arise from collisional excitation or underlying 
stellar absorption.  

In this study, we seek to derive the dust properties of I\,Zw\,18.  The presence
of dust at very low metallicities has implications both for the evolution of 
irregular and spiral galaxies, and also for cosmological models of structure 
formation in the present epoch.  It is thus imperative that we understand all 
components of such systems.   This paper is organized as follows.  In 
\S~\ref{S2} we discuss the observations and their analysis.  Presentation of 
the new \halpha\ and \hbeta\ images, and their analysis and errors, are the 
topic of \S~\ref{S3}.  In \S~\ref{S4} we discuss the locations of areas rich 
in dust, the properties of that dust distribution, and the implications for 
the molecular gas content of the galaxy which this detection affords.  We 
summarize our conclusions in \S~\ref{S5}.

\section{Observations and Data Reduction}
\label{S2}

Images were obtained with the Wide Field/Planetary Camera 2 (WFPC2) aboard 
the Hubble Space Telescope (HST) on 1998 April 4 and 1999 February 16.  In 
addition, archival WFPC2 data from programs GO-5309 and GO-5434 were 
retrieved from the STScI archive\footnote{http://archive.stsci.edu} to 
create a comprehensive dataset covering a large wavelength regime.  Here, 
we discuss the \halpha\ and \hbeta\ images, and their relevant continua from 
broad band images.  A summary of the data used is presented in Table~\ref{t1}.
Note that both \halpha\ images were obtained using the F658N filter, because 
the redshift of the emission line gas in I\,Zw\,18 \citep[\halpha\ heliocentric 
radial velocity range of 730-780 km sec$^{-1}$, or z$\sim$0.0025;][]{duf96a} 
places the 6563 {\AA} \halpha\ line in the range of peak sensitivity of the 
F658N filter.  All reductions and analysis were performed in the IRAF/STSDAS 
environment\footnote{IRAF is distributed by the by the National Optical 
Astronomy Observatories, which are operated by AURA, Inc., under cooperative 
agreement with the National Science Foundation.}.

\placetable{t1}

\subsection{Calibration and Registration}
\label{S2.1}

The data were initially processed through the standard HST pipeline.  
The pipeline-processed images were then corrected for geometric 
distortion, charge transfer efficiency, warm pixel effects, and cosmic 
ray contamination.  The background level of each image was then removed 
manually.  The archival broad band images were flux calibrated by 
multiplying by the image header keyword PHOTFLAM (erg sec$^{-1}$ cm$^{-2}$ 
{\AA}$^{-1}$), defined as the mean flux density that produces a count 
rate of 1 sec$^{-1}$, and then dividing by the header keyword EXPTIME 
(length of exposure).  The On-The-Fly Calibration scheme, implemented 
by STScI, ensures all header keywords contain the most appropriate values.  

For the narrow band images, the PHOTFLAM keyword was replaced by a synthetic 
value obtained by applying the SYNPHOT package; here, the photometric parameters 
of an observation passband are calculated individually.  Outputs which were 
used in the present analysis included the modified photometric calibration 
(output parameter URESP), the filter width, and a correction for emission 
features which appear away from the peak of a filter's sensitivity curve. 
Table~\ref{t2} contains URESP and PHOTFLAM values for narrow band images, as 
well as the system response efficiency corrections which were applied.  Finally,
all images were rotated and shifted into the orientation of the new WF3 \hbeta\ 
image.  We find our registration to be accurate to within $\sim$ 0.05\arcsec\ 
in the WF3 images (1 pixel = 0.0996\arcsec) and to within $\sim$ 0.03\arcsec\ 
in the PC images (1 pixel = 0.0455\arcsec).

\placetable{t2}

\subsection{Continuum Subtraction}
\label{S2.2}

For narrow band photometry, the removal of the underlying stellar continuum is 
a crucial component of the data analysis.  The contribution to the narrow band 
images from underlying continuum sources was found in a recursive manner.  
First, the narrow band image, which includes continuum contamination, was scaled
by a factor roughly representative of the ratio of emission contained in the 
narrow bandpass to that contained in the total wide bandpass.  Typically, this 
scaling factor was of the order of a few percent.  The resultant scaled image 
was then subtracted from the wide band image whose filter response curve most 
closely isolated the redshifted emission features in question, and which 
contained the fewest number of other significant emission lines.  Of course, 
this problem is unavoidable for certain of the emission line images, for 
example, [\ion{O}{3}] and \hbeta.  In such cases, the narrow band images were 
combined prior to the iterative subtraction process.  This procedure was 
repeated by using different scaling factors until the wide band image no longer 
retained significant diffuse structure which is indicative of emission line 
radiation.  Once this iterative procedure was completed, we were left with a 
pure continuum image, free from the line emission in question.  This scaled 
continuum image was then subtracted from the original narrow band image(s) to 
produce a pure emission line (i.e., continuum-subtracted) map.  In 
Figures~\ref{figcap1}a and \ref{figcap1}c, we show the PC \halpha\ image and 
the underlying stellar continuum which was removed from it.

The continuum scaling fraction was independently calculated using the SYNPHOT 
package.  The throughputs of the narrow band and wide filters were found at the 
wavelengths of the redshifted emission lines.  The ratio of throughputs was then
multiplied by the ratio of effective filter widths.  The fractions found by 
these two methods agree to better than one percent for both emission line 
images.  In order to obtain the total amount of line emission from I\,Zw\,18, 
the pure emission line images were then multiplied by the FWHM of each filter, 
as calculated by SYNPHOT.  We thus have an absolute calibration for the narrow 
band images.  The continuum-subtracted images in \halpha\ and \hbeta\ are 
compared in Figures~\ref{figcap1}a and \ref{figcap1}b, respectively.  The total 
\halpha\ flux from I\,Zw\,18 is found to be 3.26$\times$10$^{-13}$ erg 
sec$^{-1}$ cm$^{-2}$, in excellent agreement with the value calculated by 
\citet{dem98}, 3.3$\times$10$^{-13}$ erg sec$^{-1}$ cm$^{-2}$.

\section{Emission Line Images}
\label{S3}

\begin{figure}
\plottwo{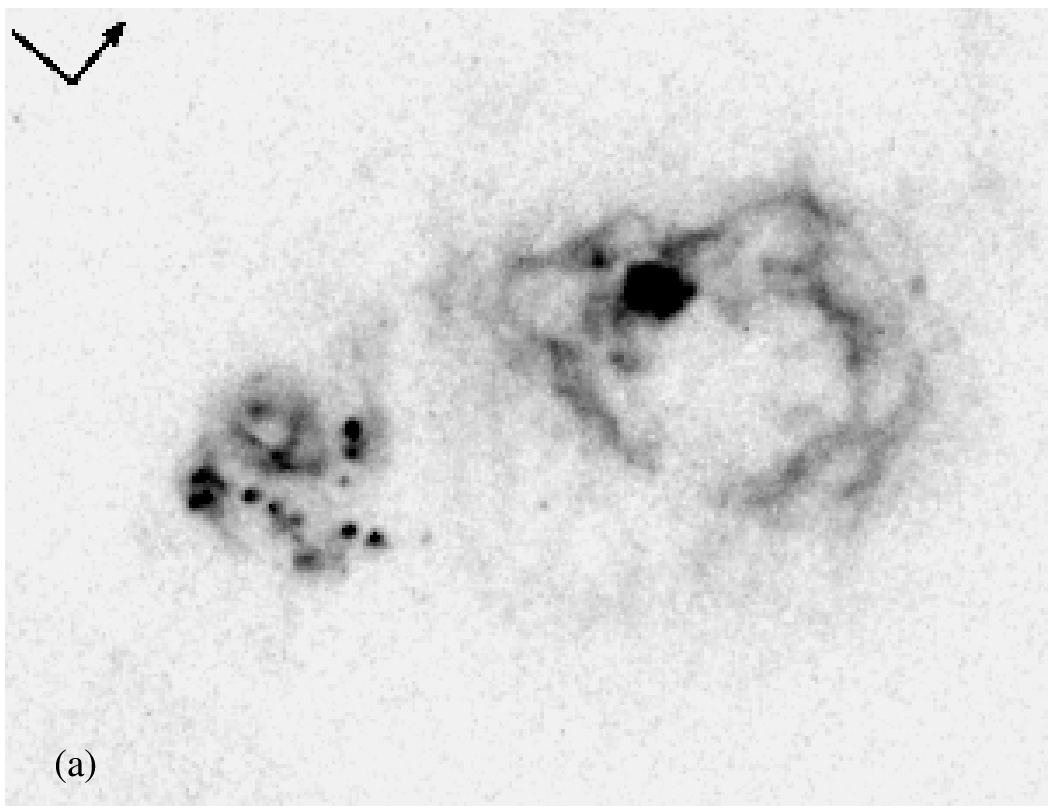}{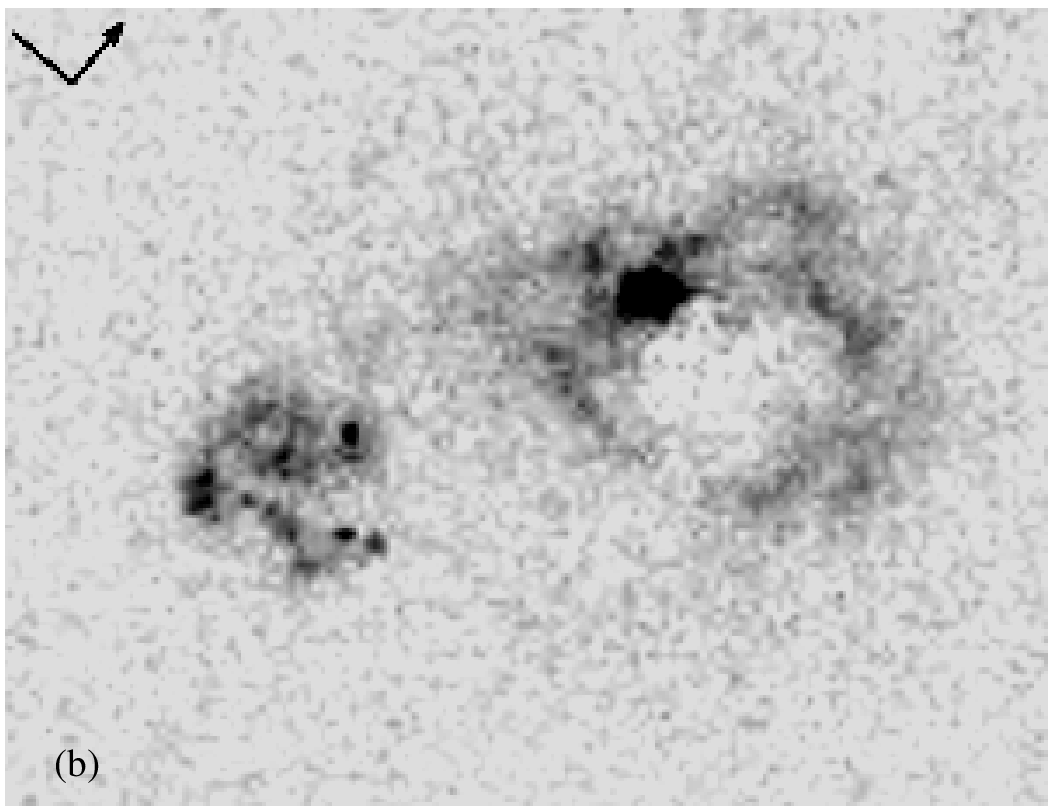}
\plotone{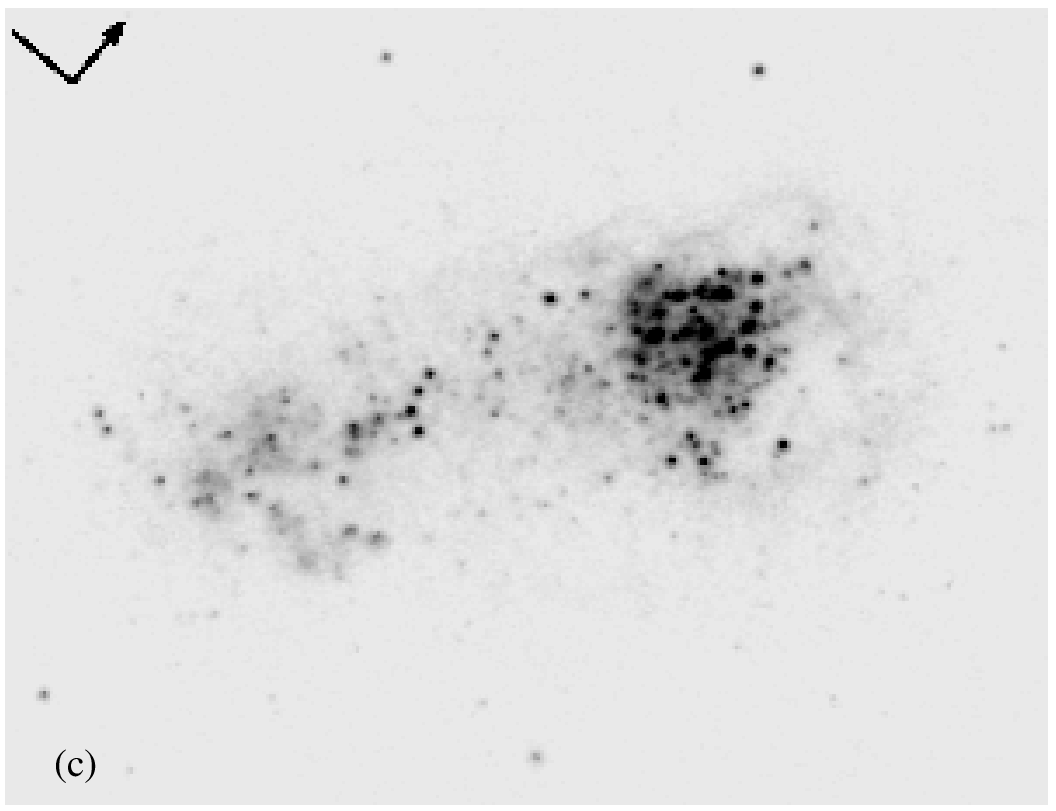}\hspace{0.45 cm}
\plotone{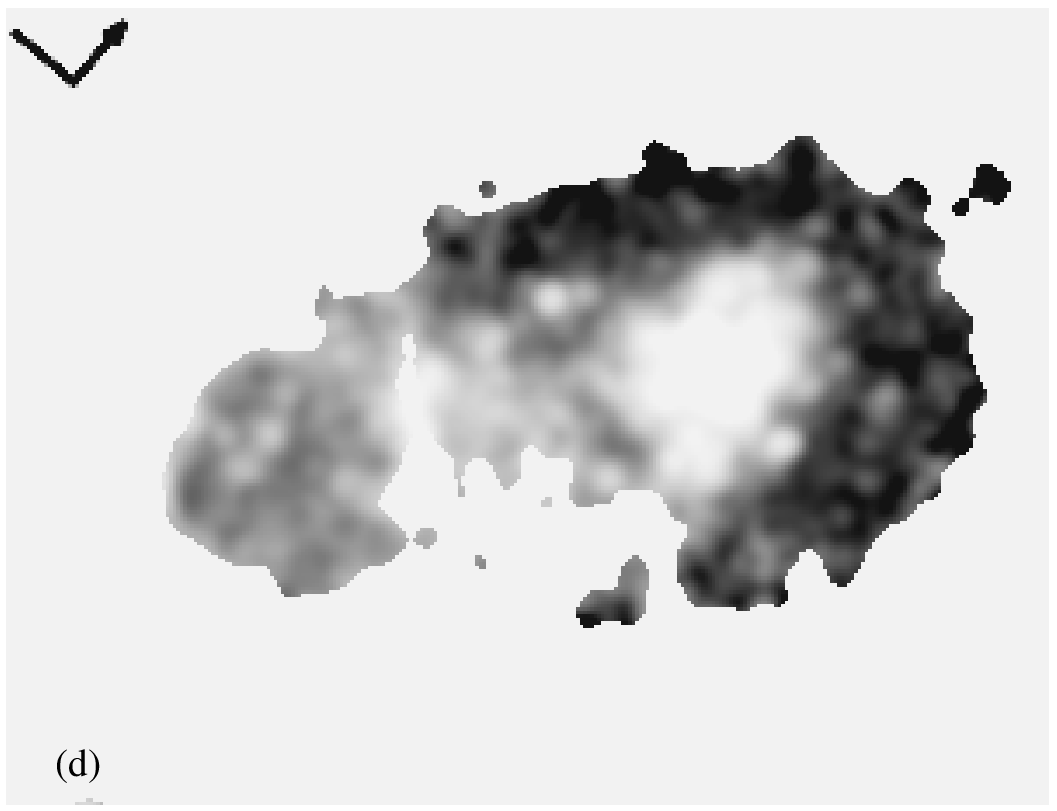}
\caption{Continuum-subtracted \halpha\ (a) and \hbeta\ (b) pure emission line 
images of I\,Zw\,18, compared with the scaled emission line subtracted R-band 
image (c) and the (smoothed) equivalent width of the \halpha\ emission (d).  The
scale encompasses the full range of intensity for all images, except for (d), 
where the scale is from 100 \AA\ or less (white) to 2000 \AA\ or greater 
(black).  The field of view, identical in all images, is 13.7$\arcsec$ $\times$ 
10.5$\arcsec$, or 833 pc $\times$ 641 pc at the adopted distance of I\,Zw\,18.  
The arrows denote north (tipped) and east and are 1$\arcsec$ in length.  The 
lower signal-to-noise ratio and resolution of the \hbeta\ image compared to the 
\halpha\ image will form the dominant contribution to the errors in the \hab\ 
flux ratio.  Note the negligible residual ionized gas in the stellar continuum, 
which is slightly displaced from the ionized gas; there are also stars which are
outside the luminous extent of I\,Zw\,18 in \halpha.  The \halpha\ equivalent 
width map (d) was smoothed to $\sim$ 0.25\arcsec\ resolution, and was masked to 
only retain information in areas of high signal-to-noise ratio in the original 
\halpha\ image (for clarity).  A comparison of (d) with the aperture placements 
used to derive the \hab\ flux ratios (see Figure~\ref{figcap3}a, and discussion 
in Sections~\ref{S3.1} and \ref{S4.1.1}) shows that all apertures encompass 
areas with large equivalent width ($>$ 700 \AA\ in \halpha\ and $>$ 80 \AA\ in 
\hbeta; see Table~\ref{t4}), making negligible the potential effects of stellar 
absorption.}
\label{figcap1}
\end{figure}

\begin{figure}
\plotone{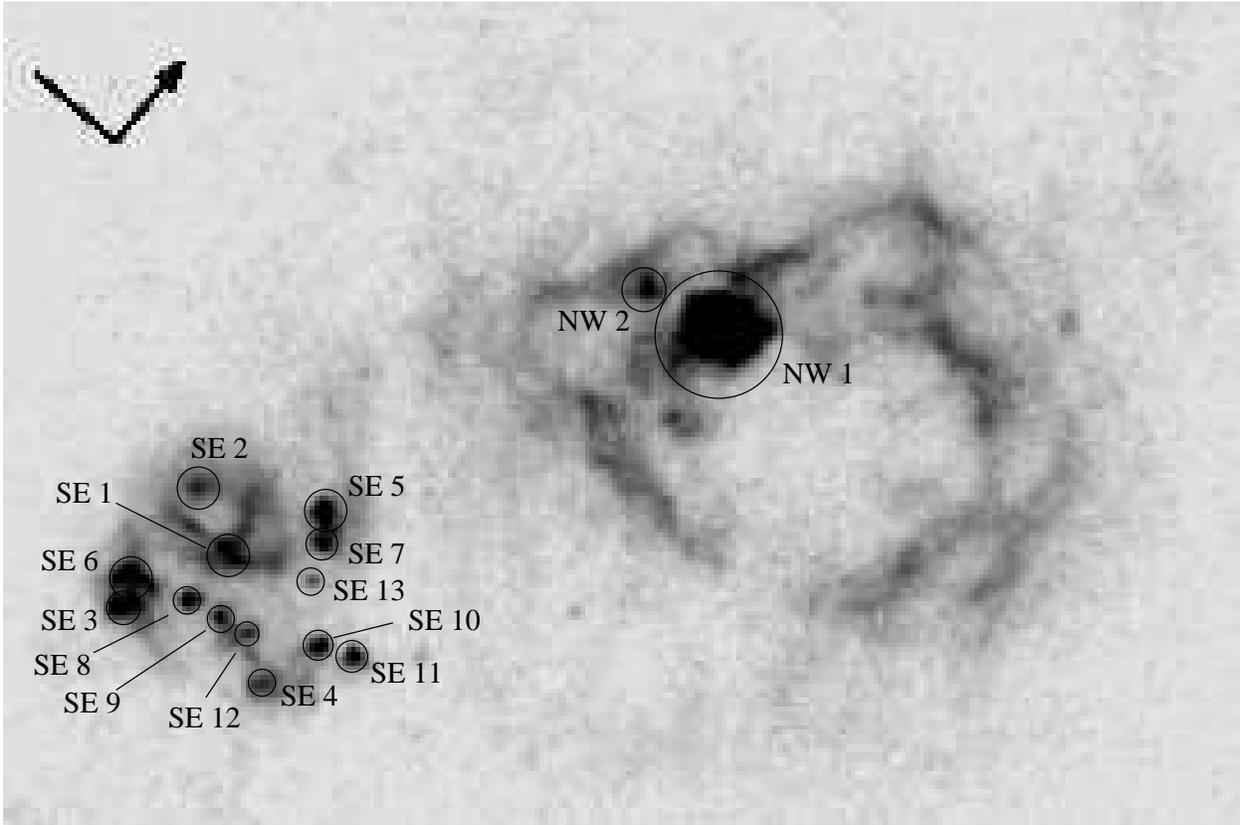}
\caption{The bright \HII\ regions of I\,Zw\,18, overlaid on the 
continuum-subtracted PC \halpha\ image.  The characteristics of the regions, 
numbered in order of decreasing \halpha\ luminosity,  are summarized in 
Table~\ref{t3}.}
\label{figcap2}
\end{figure}

The ionized gas distribution of I\,Zw\,18 is well-resolved in the PC \halpha\ 
image (Figure~\ref{figcap1}a).  A wispy, filamentary structure with prominent 
shells dominates the morphology of the galaxy in the light of \halpha.  The very
bright \halpha\ concentration in the NW region is slightly offset from the peak 
of continuum emission (compare Figures~\ref{figcap1}a and \ref{figcap1}c).  The 
Southeast (SE) region of the galaxy shows more bright knots in \halpha\ than 
does the NW region, and less filamentary structure.  We do detect a smaller 
shell in the SE region, however; as will be shown in \S~\ref{S4}, this area is 
also coincident with a prominent dust concentration.  The relative shell sizes 
in the NW and SE regions and the presence of more OB associations and less 
extended diffuse emission in the SE region is in good agreement with the 
observations by HT95\nocite{hun95} and \citet{duf96b} that this region contains 
younger stars than the NW component.  The bright \HII\ regions in I\,Zw\,18 as 
revealed in the PC image are shown in Figure~\ref{figcap2} and detailed in 
Table~\ref{t3} in order of decreasing \halpha\ luminosity.  Our catalog compares
well with that of HT95\nocite{hun95}; all of our features are also detected and 
cataloged therein, except for SE 13, the least-luminous \HII\ region detected, 
which was not visible in their F656N filter \halpha\ image.

\placetable{t3}

The continuum-subtracted \hbeta\ image (Figure~\ref{figcap1}b) shows a very 
similar structure to the \halpha\ image.  The signal-to-noise here is lower than
in the \halpha\ image (due to lower system throughput than the F658N filter 
(roughly a factor of 2 less), a shorter integration time, and the intrinsically 
fainter \hbeta\ emission line strength), and produces the limiting uncertainty 
in the calculated reddening values across the galaxy (see below).  Note the 
general agreement in morphology of the two images; every filament of emission in
\halpha\ is also detected in \hbeta, down to the limit of the resolution in the 
latter image.  Resolution of fainter \HII\ regions is lost in the \hbeta\ image 
(e.g., SE 13; see Figures~\ref{figcap1}b \& \ref{figcap2}, and Table~\ref{t3}).  

\subsection{Reddening in I\,Zw\,18}
\label{S3.1}

Regions with high photon counts (i.e., large equivalent widths, hereafter 
denoted by \eqwhalpha\ and \eqwhbeta\ for \halpha\ and \hbeta, respectively) in 
both \halpha\
and \hbeta\ were next analyzed using aperture photometry to determine fluxes.  
Apertures were placed on identical regions in six different images: \halpha\ and
\hbeta\ fluxes, \halpha\ and \hbeta\ photon counts, and continuum images used 
for each filter.  To make sure that we were obtaining proper flux calibration of
the \halpha\ images, we performed this procedure for both the WF3 and PC F658N 
filter images, with continuum images selected to match the original resolution 
(i.e., PC continuum image used for PC \halpha\ continuum subtraction, and 
likewise for the WF3 image).  Table~\ref{t2} (last two columns) contains data 
on the continuum subtraction values used to determine aperture flux ratios.

\begin{figure}
\plottwo{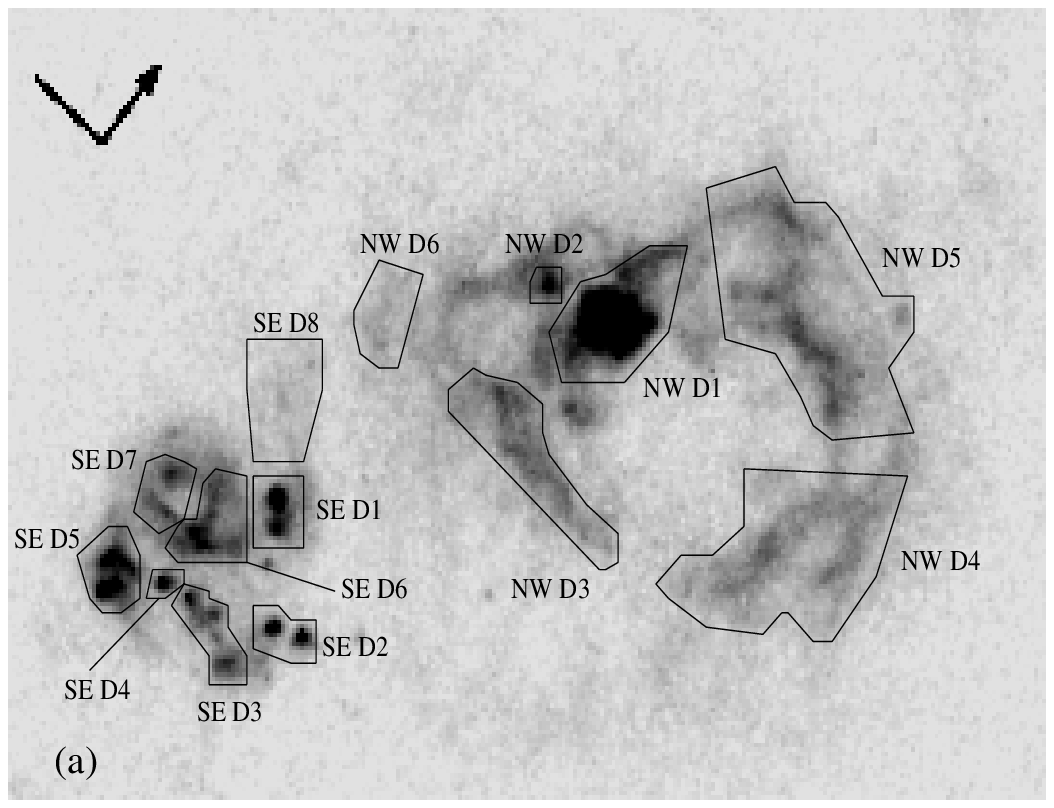}{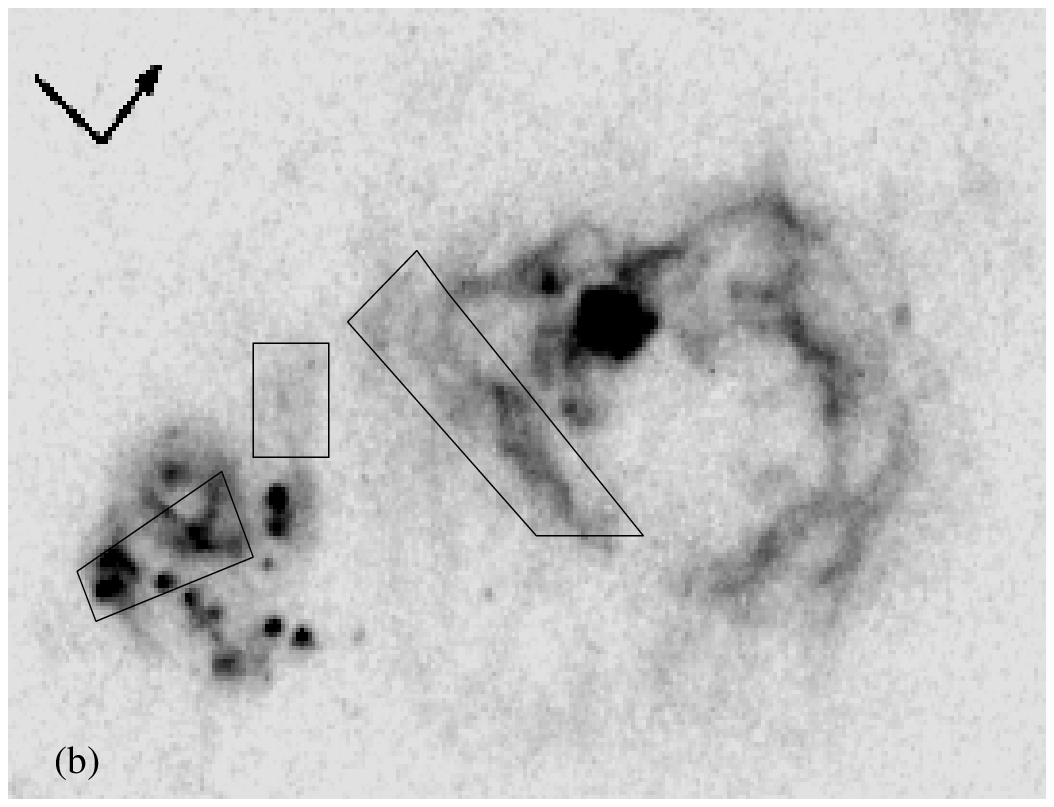}
\caption{Continuum-subtracted \halpha\ image of I\,Zw\,18, overlaid with the 
locations of apertures used to determine \hab\ flux ratios (a).  The apertures 
are labeled, and the reddening and extinction to each is shown in 
Table~\ref{t4}.  In (b), the same image is overlaid with the locations of 
significant dust concentrations (i.e., \hab\ $\geq$ 3.0).  Some of the dust-rich
regions are near ionizing sources (e.g., in the SE region), while others are 
displaced from clear sources of UV flux.  These dust concentrations may isolate 
areas which are rich in molecular gas; note that these are not necessarily 
coincident with sightlines to UV-bright clusters \citep[c.f.,][]{vid00}.}  
\label{figcap3}
\end{figure}

Upon preliminary analysis of the data, we found that using small circular 
apertures centered on the \HII\ regions (i.e., the apertures of the \HII\ 
regions as shown in Figure~\ref{figcap2} and listed in Table~\ref{t3}) yielded 
unacceptably large errors for most of the catalogued \HII\ regions; the sources 
of these errors are discussed below.  For this reason, we chose to expand the 
apertures to larger polygons which included the majority of visible flux when 
overlaid on each image, encircled areas of only high \eqwhalpha\ and \eqwhbeta\ 
(compare Figures~\ref{figcap3}a and \ref{figcap1}d), and often contained 
multiple \HII\ regions in a single aperture.  This technique minimizes potential
errors caused by slight mis-registration of the images, and enhances the 
signal-to-noise ratio by increasing photon counts in each aperture (see below 
for further discussion).  The locations of these larger polygonal apertures are 
shown in Figure~\ref{figcap3}a.

Flux ratios were then calculated by dividing \halpha\ fluxes by \hbeta\ fluxes 
in each aperture.  In the absence of dust, the \hab\ ratio is confined to a 
narrow range of values that is relatively insensitive to temperature and 
density.  Following SK93\nocite{ski93}, we adopt an electron temperature of 
19,000 K for I\,Zw\,18; we thus expect \hab\ = 2.76, based on the Balmer line 
ratios from \citet{hum87}.  Values that are elevated above this theoretical 
ratio are indicative of the presence of dust, which preferentially absorbs and 
scatters shorter wavelength radiation.  We list in Table~\ref{t4} the applied 
apertures, \halpha\ and \hbeta\ fluxes, and flux ratios with errors.  We note 
that all \hab\ ratios are consistent with the expected minimum theoretical 
value, and that the small errors make the detections of dust significant.

\placetable{t4}

\subsection{Error Sources} 
\label{S3.2}

A potential major source of uncertainty in extragalactic reddening studies is 
the amount of foreground extinction due to gas and dust within the Galaxy.  At a
Galactic latitude of $+$45\arcdeg, foreground extinction is expected to be low 
in I\,Zw\,18.  In the direction of I\,Zw\,18, the (0\fdg3$\times$0\fdg6 
resolution) maps of \citet{bur82} indicate a foreground extinction of E(B-V) 
$\sim$ 0.01 magnitudes.  The more recent, higher resolution ($\sim$ 6$^{\prime}$
FWHM) study of Schlegel, Finkbeiner, \& Davis (1998)\nocite{sch98} indicates 
that a marginally higher value of E(B-V) $\sim$ 0.03 magnitudes is applicable.  
Inspection of the extinction maps of this latter study indicates that there is no
appreciable gradient in or fluctuation of E(B-V) across the luminous extent of 
I\,Zw\,18; the average value applies over scales much larger than the field of 
view under study here (e.g., Figures~\ref{figcap1}(a)-(d)), and the foreground 
extinction is smooth on these scales.  

For the purposes of this study, we seek to constrain the amount of internal 
extinction within I\,Zw\,18.  While there is a moderate amount of extinction 
between the observer and the galaxy, which we correct for in the following 
analysis, any variations in the \hab\ flux ratio observed are likely to be 
intrinsic to I\,Zw\,18.  Although we cannot rule out small-scale (i.e., 
arcsecond scales) local maxima in the foreground extinction which may affect our
interpretation, we favor the hypothesis that searching for variations in the 
\hab\ flux ratio allows us to constrain the nature of dust within I\,Zw\,18 
itself.  The studies by \citet{bur82} and \citet{sch98} are consistent with 
values of 0 $<$ E(B-V) $<$ 0.03; we adopt an intermediate value of E(B-V) $=$
 0.02.  Assuming R$_V$ $=$ 3.1, we then have a foreground extinction of 0.06 
magnitudes, which is small compared to our regions of highest detected 
extinction.  Following both of the aforementioned studies, we attach to this 
foreground value a 25\% uncertainty, and propagate this through our final 
extinction values calculated in Section~\ref{S4.1.1}.

Uncertainties were calculated for the observed reddening values by applying 
Poisson statistics.  We discuss other potential sources of error below; however,
this should represent an accurate measure of the final error in the \hab\ ratio 
due to the lower signal-to-noise ratio of the \hbeta\ WF3 image compared to the 
\halpha\ image.  This dominant error term limits which sections of the image, 
and hence what physical conditions, may be sampled for the presence of dust.  
For example, the central region of the NW shell cannot be tested for the 
presence of dust, due to the fact that the number of detected \hbeta\ photons is 
too small to allow physically meaningful errors to be assigned.  

Other sources of error in the flux calibration and image analysis procedures are 
discussed and deemed negligible as follows.  We implicitly assume that removal 
of DC levels from images (e.g., background removal, scaling of images by 
numerical fractions) does not affect the standard deviation of the image, and 
therefore does not affect the error terms derived.  Errors in continuum 
subtraction fractions were assessed by performing photometry on images with 
different continuum fractions than the final ones used.  Changing the continuum 
subtraction fractions is a second order effect compared to the errors in number 
of counts.  The values which we apply for continuum subtraction are also 
supported by the fact that the derived \hab\ flux ratios are all consistent with
the lower limit theoretical ratio of 2.76 \citep[][and arguments below]{hum87}.  
Furthermore, inspection of the continuum images (after removal of the 
narrow band, ionized component) reveals very little ionized gas remaining, 
indicating successful continuum subtraction (see Figures~\ref{figcap1}a and 
\ref{figcap1}c).

We assume that the position angle of the WFPC2 camera is exactly known.  
Rotation of the images into the orientation of the new WF3 images was 
accomplished using simple linear interpolation between pixels.  This method was 
favored due to its simplicity; we found no reason to apply more complicated 
interpolation schemes.  Shifting the rotated images to align with the narrow 
band images is not an error-free process; as stated previously, we expect that 
image registration is accurate only to within a fraction of a pixel.  Using the 
aforementioned larger polygonal apertures (see Figure~\ref{figcap3}a), we obtain
higher photon counting statistics, ensuring that the contribution to the final 
error terms due to image rotation and mis-registration are of second order or 
higher in all cases.

\section{Dust In I\,Zw\,18}
\label{S4}

In \S~\ref{S3}, we unambiguously detect elevated \hab\ flux ratios in various 
areas inside I\,Zw\,18 (see Table~\ref{t4}).  In this section, we characterize 
the dust content and denote some implications for the evolutionary status and 
molecular gas content of I\,Zw\,18.  

\subsection{Dust Properties}
\subsubsection{Extinction, Errors And Potential Systematic Errors}
\label{S4.1.1}

We calculate reddening values for each aperture discussed in \S~\ref{S3} by 
applying the relation
\begin{equation}
\frac{I_{\lambda}}{I_{H\beta}} = \frac{I_{{\lambda}_0}}{I_{{H\beta}_0}} 
10^{-c\; [f({\lambda}) - f({H\beta})]}
\label{e2}
\end{equation}
\noindent where \begin{math}I_{\lambda}/I_{H\beta}\end{math} is the observed 
\hab\ flux ratio, \begin{math}I_{{\lambda}_0}/I_{{H\beta}_0}\end{math} is the 
theoretical ratio \citep[2.76,][]{hum87} for the adopted temperature of 19,000 K
(SK93)\nocite{ski93}, c is the logarithmic reddening correction at \hbeta, and 
\begin{math}f(\lambda) - f(H\beta)\end{math} (= -0.37) is calculated from the 
interstellar extinction curve of \citet{sea79} as specified by \citet{how83}.  
We then find c and solve for the extinction in magnitudes by applying the 
conversion factor between c and A$_{\rm V}$ of \citet{sch77}, where A$_{\rm V}$ 
$=$ 2.17~$\cdot$~c.  Finally, we correct for 0.06 magnitudes of foreground 
extinction \citep{bur82,sch98} as discussed in \S~\ref{S3.2}.  These values are 
also listed in Table~\ref{t4}; note that all extinctions are rather small, the 
largest being $\sim$ 0.5 magnitudes, and a fiducial value being 0.2 magnitudes 
for the SE region.  Thus, large spatial variations in internal extinction in 
I\,Zw\,18 are not sufficient to compromise previous nebular abundance studies; 
however, the statistically significant elevation of some \hab\ flux ratios 
indicates that appreciable amounts of dust do exist within the galaxy and merit 
further investigation.

A potential difficulty in interpreting elevated \hab\ ratios as indicative of 
the presence of dust is the contribution from collisional excitation of \HI.  
\citet{dav85} and SK93\nocite{ski93} have discussed the possibility that the 
high electron temperatures in I\,Zw\,18 may lead to collisional excitation of 
the neutral hydrogen, which would affect the emission line ratios of some of the
lower-level hydrogen transitions, most notably \halpha\ and \hbeta.  Since this 
effect is directly proportional to the neutral hydrogen fraction, an 
observationally challenging quantity to measure, the importance of this effect 
has been difficult to assess.  Photoionization models by \citet{sta99} suggest 
that collisional excitation effects may affect the \hab\ ratio by as much as 
11\% in the NW region of I\,Zw\,18.  The distribution of enhanced \hab\ ratios 
observed here argues against the importance of collisional excitation of the 
neutral hydrogen.  Most of the elevated \hab\ ratios are seen in the SE 
component, which has a lower electron temperature than the NW region 
\citep[SK93\nocite{ski93};][]{izo98}.  Furthermore, there is no evidence of 
enhanced \hab\ flux ratios in areas of expected lower ionization parameter (i.e.,
radial gradients around ionizing stars).  Thus, we interpret the regions which 
demonstrate elevated \hab\ ratios (see further discussion below) to be regions 
with enhanced extinction.  While we cannot rule out the possibility that 
collisional processes are making a limited contribution to the \halpha\ emission
in areas near UV-bright clusters (e.g., sections in the SE region; see 
Figure~\ref{figcap3}a), the effect is minimal even in these regions.  For 
example, apertures SE D3 and SE D4 (see Table~\ref{t4} and 
Figure~\ref{figcap3}a) are spatially separated by only 21.7 pc.  The 
statistically different \hab\ flux ratio between the two suggests that, if 
collisional excitation is taking place, its effects are not severe enough to 
compromise the interpretation of the photometry of this study.

A second effect which may elevate \hab\ ratios above the theoretical value is an 
error in the assumed electron temperature, \te.  If \te\ is much lower than the 
adopted value, any elevated \hab\ ratios will be caused by the preferential 
excitation of the lower-energy \halpha\ transition compared to \hbeta.  
SK93\nocite{ski93} find \te\ to be lower in the SE region than in the NW (17,200
$\pm$ 1,200 K and 19,600 $\pm$ 600 K, respectively).  However, as will be shown 
below, we find \hab\ flux ratios (corrected for foreground extinction) which are
as high as 3.4.  To attain such ratios as a consequence of an overestimated \te,
the actual temperature would have to be as low as 2,500 K in some areas, and 
would have to vary on unphysically short spatial scales (of order parsecs).  
Neither of these effects are observed, nor are they predicted by models of the 
photoionization structure of the galaxy \citep{sta99}.  We thus conclude that 
errors in the adopted \te\ provide a negligible error contribution in the 
following analysis.

Finally, it is possible that underlying stellar absorption may affect the pure 
emission line images.  In regions with low \eqwhalpha\ and \eqwhbeta, underlying
stellar Balmer absorption may lower the \hbeta\ emission line flux relative to 
\halpha, simulating the effect of dust.  To look for this effect we have 
tabulated the \eqwhbeta\ and \eqwhalpha\ for each of our measured regions; 
Figure~\ref{figcap1}d shows the smoothed \eqwhalpha\ map.  If underlying 
absorption were important, one would expect an anti-correlation between the 
measured extinction and the \eqwhbeta.  In fact, the data in Table~\ref{t4} do 
show weak evidence of this trend.  A regression of A$_V$ on \eqwhbeta\ yields a 
correlation coefficient of -0.5.  This is not direct evidence that some of the 
enhancement in the \hab\ ratio is due to underlying stellar absorption (it may 
be that dust is found preferentially associated with lower surface brightness 
stellar populations), but the effect does need to be considered.  From the 
modeling of \citet{gon99}, the \eqwhbeta\ of an instantaneous burst rises 
monotonically from $\sim$ 3 \AA\ to $\sim$ 10 \AA\ for ages of 0 to 200 Myr.  In
a constant star formation model \eqwhbeta\ rises monotonically from $\sim$ 3 
\AA\ to $\sim$ 6 \AA\ over the same period.  If we could be confident that all 
of the stars in I\,Zw\,18 were younger than 20 Myr, then an upper limit of 6 
\AA\ \eqwhbeta\ of underlying \hbeta\ absorption would be reasonable.  If there 
are stars older than this in I\,Zw\,18, then the continuous star formation model 
would be more appropriate (which has a maximum value of $\sim$ 8 \AA\ \eqwhbeta\
for ages of $\sim$ 0.5 Gyr and higher).  For our lowest \eqwhbeta\ regions 
($\sim$75 to $\sim$100 \AA), some, if not most, of the enhancement in the \hab\ 
ratio could be due to underlying stellar absorption if it were mixed with an 
intermediate age stellar population.  However, for the regions with A$_V$ in 
excess of 0.15 (\hab\ $=$ 3.00) it is unlikely that the enhancement in the \hab\
ratio could be due solely to underlying stellar absorption.  Note that in 
several regions (with \eqwhbeta\ $=$ 120 to 160 \AA, see Table~\ref{t4}) the 
\hab\ ratio is consistent with no reddening (and thus no effect of underlying 
absorption).  Thus, while we cannot completely rule out some enhancement in the 
\hab\ ratio due to underlying stellar absorption, we can rule out that it is the
dominant effect in the regions where we have the highest \hab\ ratios.   Since 
these regions are the focus of what follows, we will assume that the effects of 
underlying absorption are negligible. 

\subsubsection{Dust Locations And Masses}
\label{S4.1.2}

In Figure~\ref{figcap3}b, we overlay apertures on the PC \halpha\ image of 
I\,Zw\,18 which outline the locations of elevated \hab\ ratios (note that these 
trapezoidal apertures are meant to guide the eye to areas with \hab\ $\geq$ 3.0,
and in some cases contain more than one of the apertures used to derive the 
\hab\ flux ratios).  We detect dust in regions which are near UV-bright sources,
as well as in regions which are far away from them.  The high surface brightness
\HII\ regions which have significant extinction are all located in the SE 
component, in agreement with the ground based optical observations 
\citep[SK93\nocite{ski93};][]{izo98,vil98}.  This is expected, since the SE 
region contains younger stars and likely more dense molecular clouds than the NW
region.  The distribution of dust (and therefore molecular gas) suggests that 
future star formation is likely in the SE region; furthermore, the reddening 
morphology hints at the presence of a band of extinction between the SE and NW 
components.  This region cannot be sampled for dust with the present \hab\ ratio
method, due to the absence of \halpha\ and \hbeta\ emission in this region (see 
\S~\ref{S3.2}).  Overall, we detect dust with a similar distribution to that 
found in the dwarf starburst galaxy NGC 5253 \citep{cal97}.  The irregular 
nature of the dust distribution immediately implies one or both of the following 
scenarios.  First, dust which is near UV-bright clusters is surviving in a very 
hard radiation field, implying an efficient mechanism of self-shielding on small
scales and the prominence of clumpy structure in the ISM.  Second, dust which is
displaced from UV-bright clusters is consistent with dust production by a 
previous generation of star formation.

In order to estimate a mass of the dust which we detect, we must make certain 
assumptions about the nature of the grains.  One method with which to obtain a 
dust mass estimate is to use the ratio of intensities of far infrared (FIR) 
emission at 60 and 100 $\mu$m.  However, I\,Zw\,18 is not detected by the IRAS 
satellite in either waveband.  A second method with which to find the dust mass 
is to apply a detailed balance argument, assuming one knows the nature of the 
incident radiation field and the composition of the grains.  However, especially
in the case of BCD's such as I\,Zw\,18 where the nature of the underlying 
stellar population is not yet fully characterized, this detailed balance 
argument is also not applicable.

A third and widely used method to obtain a dust mass is to apply a gas-to-dust
ratio which is applicable to similar physical conditions (metallicity, radiation
field, star formation efficiency, etc.).  Seminal Galactic work on this 
correlation by Bohlin, Savage, \& Drake (1978)\nocite{boh78} finds that the mean
ratio $<$(\HI\ $+$ H$_2$)/E(B-V)$>$ is nearly constant (within a factor of 1.5),
with a value of 5.8$\times$10$^{21}$ cm$^{-2}$ mag$^{-1}$.  From a larger 
stellar sample, \citet{dip94} find a slightly lower value of 
4.93$\times$10$^{21}$ cm$^{-2}$ mag$^{-1}$.  Extinction studies in other 
environments suggest a trend of increasing gas-to-dust ratio with decreasing 
ambient metallicity.  Extinction studies toward stars in the LMC arrive at an 
average value of $<$(N$_{\rm H~I}$)/E(B-V)$>$ = 2.4$\times$10$^{22}$ cm$^{-2}$ 
mag$^{-1}$ \citep{fit85,duf82,cla96}.  The more metal-poor conditions of the SMC
yield yet higher values, perhaps as high as 10$^{23}$ cm$^{-2}$ mag$^{-1}$ 
\citep{pre84}.

The \HI\ synthesis imaging study of \citet{van98} shows that there is a column 
density maximum of 3$\times$10$^{21}$ cm$^{-2}$ which is coincident with the SE 
component (see their Figure 6).  If we assume that the SE ionized component is 
embedded in an \HI\ envelope, then we can make a very crude estimate of the 
ratio of \HI\ column density to extinction by applying the relation  
\begin{equation}
\frac{\rm N(H~I)}{\rm E(B-V)} \sim \frac{1.5\times10^{21} \rm\; cm^{-2}}{0.06 
\rm\; mag} 
\sim 2.3\times10^{22} \rm\; cm^{-2} \rm\; mag^{-1}
\label{e3}
\end{equation}
\noindent where we have assumed R$_{\rm V}$ = A$_{\rm V}$/E(B-V) = 3.1, and that
a representative value of the average extinction towards the SE region is 
A$_{\rm V}$ = 0.2 mag.  This value is comparable to, but less than, estimates of
the gas-to-dust ratio in the SMC \citep[3.7-5.2$\times$10$^{22}$ cm$^{-2}$ 
mag$^{-1}$;][]{bou85}.  However, it makes no correction for a potentially large 
contribution from molecular hydrogen (necessary to sustain the high star 
formation rate; see arguments below), which may allow the two results to be 
consistent.  Furthermore, this estimate only accounts for the fraction of the 
dust which may be foreground to I\,Zw\,18; if a substantial amount of dust is 
located behind or mixed within the \HII\ regions, this argument will (perhaps 
significantly) underestimate the total dust mass within the galaxy.  

Next, we must assume a value for the dust-to-gas ratio by mass in order to 
obtain a dust mass estimate.  We make such an assumption in two ways.  First, we
note that \citet{sta00} finds the average dust-to-gas ratio in the SMC to be 
8.2$\times$10$^{-5}$ by mass (roughly 30 times less than in the Galaxy).  Since 
this is the lowest-metallicity galaxy which has well studied extinction 
properties, we naively apply this ratio to I\,Zw\,18.  This represents an 
apparent upper limit, since there is evidence of a trend towards lower 
dust-to-gas ratios at lower metallicities, as previously mentioned, and the 
metallicity of I\,Zw\,18 is $\sim$ 0.8 dex lower than that of the SMC.  Applying
such an assumption and Equation~\ref{e3}, rewritten in the more convenient form
\begin{equation}
A_{\rm V}/N_{\rm H} = 1.6\times10^{-22} \; \rm{mag \; cm^2}
\label{e4} 
\end{equation}
\noindent we arrive at a dust mass of $\sim$ 2000 \msun\ (with no correction for
the molecular gas content).  

The second method with which to estimate the dust-to-gas ratio is to apply the 
analytical relation found by \citet[][hereafter LF98]{lis98}, \begin{math}12 + 
\rm{log(O/H)} \propto (0.52\pm0.25) \cdot log(M_{dust}/M_{HI})\end{math}.  While
this relation was derived for dwarf irregulars and not BCD's, we include it for 
completeness, and because the application of the relation to the SMC yields a 
dust-to-gas ratio of 6.5$\times$10$^{-5}$ by mass, in general agreement with the
observational result of \citet{sta00}.  Applying the same relation to I\,Zw\,18,
we obtain a dust-to-gas ratio of 1.7$\times$10$^{-4}$ by mass, which is 
consistent with the upper limit of 3.7$\times$10$^{-4}$ obtained by 
LF98\nocite{lis98} by assuming a 100 $\mu$m flux for I\,Zw\,18.  We thus assume 
a value of 2$\times$10$^{-4}$ for the dust-to-gas ratio using this relation, but
note its general uncertainty, and that it appears high in comparison to 
observational estimates from other low-metallicity systems 
\citep[e.g.,][]{sta00}.  Application of Equation~\ref{e4} then yields a dust 
mass of $\sim$ 4800 \msun.

From the above arguments, we estimate a total dust mass of (2-5)$\times$10$^3$ 
\msun\ within I\,Zw\,18, due to the lack of further constraints which may be 
placed on the nature of the dust content from these observations.  We note that 
this dust mass is lower than all of the BCD's which are detected in the IRAS 
bands in the study by LF98\nocite{lis98}, as expected for the very low 
metallicity of I\,Zw\,18.  We thus consider this a conservative estimate of the 
total dust content of this BCD.  If the dust production process and evolution 
behaves as in the dwarf irregular sample (e.g., all dwarf irregulars with 12 + 
log(O/H) $<$ 7.4 have dust masses $<$ 10$^3$ \msun; LF98\nocite{lis98}), then 
the above values may be too high.  Further constraints on the nature of the dust
content of I\,Zw\,18 will be very useful. 

\subsection{Implications For The Molecular Gas Content}
\label{S4.2}

I\,Zw\,18 is undergoing an intense star formation episode, as evidenced by its 
blue photometric color \citep[U$-$B = -0.88, B$-$V = -0.03;][]{van98}, its large
number of ionizing, massive stars (see Table~\ref{t3}), and by the large ionized
gas extent (e.g., Figures~\ref{figcap1}a and \ref{figcap1}b).  Consequently, we 
expect concentrations of molecular gas throughout the galaxy, since we observe 
large \HII\ regions and OB associations.  I\,Zw\,18 is not detected in CO 
observations, nor is any galaxy with ambient metallicity below 12 + log(O/H) 
$\sim$ 8.0 \citep{tay98,bar00}.  The potential reasons why detections of CO in 
low-metallicity galaxies have remained elusive are numerous.  One explanation is
that the conversion factor, $X_{\rm CO}$, from I$_{\rm CO}$ to N$_{\rm H_2}$ 
becomes nonlinear below a certain metallicity \citep{mal88}.  At low abundances,
the column density of the CO molecule will decrease.  In areas undergoing 
massive star formation, the hard UV radiation field will more easily 
photodissociate CO to the point where it can no longer be self-shielding; 
\citet{mad97} find observational evidence for such behavior.  This will decrease
the component of molecular clouds which will have prominent CO emission, while 
the H$_2$ is less affected.  The filling factor decreases, reducing the CO 
luminosity for a given molecular gas mass, and incompletely sampling the H$_2$ 
mass as a result. 

These large dust concentrations (of order 50-100 pc, see Figure~\ref{figcap3}b)
suggest that the process of H$_2$ formation on grain surfaces may be ongoing in 
I\,Zw\,18.  Recent FUSE non-detections of H$_2$ in absorption in I\,Zw\,18 
\citep{vid00} can still be consistent with the presence of large amounts of 
molecular gas, since these HST results indicate that at least some of the dust 
is peripheral to the UV-bright clusters, and not necessarily along lines of 
sight to them.  We interpret these results as evidence for a clumpy ISM in 
I\,Zw\,18; small, dense molecular clouds will allow active star formation, while
their volume filling factor and overall CO luminosity will be very small.  
\citet{vid00} also conclude that dense, discrete clouds may have gone undetected
in their FUSE observations.  Clearly, very high-resolution observations 
\citep[i.e., interferometric CO mapping; e.g.,][]{wal01} of the ISM of I\,Zw\,18
and of other low-metallicity galaxies are needed to attempt to locate the 
molecular gas.

It is most intriguing that the two lowest-metallicity galaxies known, I\,Zw\,18 
and SBS\,0335-052 \citep[12+log(O/H) = 7.33$\pm$0.01,][]{izo97}, both contain
significant amounts of dust.  As discussed above, in I\,Zw\,18, some of the dust 
concentrations are displaced from ionized regions.  Similarly, in SBS\,0335-052,
dust is found displaced from the locations of the six massive star clusters 
\citep{thu97}.  In both galaxies, the presence of dust displaced from some of 
the ionizing clusters suggests that the molecular gas content of low-metallicity
galaxies is very difficult to quantify via UV absorption or CO tracer lines.  In
particular, these results suggest that the lack of molecular gas as inferred 
from such investigations does not necessarily imply a lack of molecular gas, but
rather a different mechanism of shielding (e.g., small, dense molecular clouds) 
when the metallicity is very low.  

\section{Conclusion}
\label{S5}

New WFPC2 narrow band images of the blue compact dwarf galaxy I\,Zw\,18 have been
analyzed.  Images at \halpha\ and \hbeta\ are used to constrain the dust content
and distribution of the galaxy, to analyze the morphology of the ionized gas, 
and to characterize individual \HII\ regions.  We detect significantly elevated 
\hab\ flux ratios throughout the galaxy, both close to and displaced from UV 
sources, and in general agreement with the morphology derived by \citet{vil98}. 
The effects of collisional excitation and stellar absorption are shown to be 
minimal, and we interpret these ratios as indicative of the presence of dust.  
The small total dust mass, estimated to be of order 10$^3$ \msun, is expected 
for the very low metallicity of the galaxy and is consistent with the IRAS 
non-detection in the far infrared regime.  The presence of dust at low 
metallicities is now seen in the two lowest metallicity galaxies and is 
significant, since these may be the most ideal testbeds for studies of galaxy
formation and chemical evolution processes at low redshifts.

The presence of dust in I\,Zw\,18 allows us to form a consistent picture of the 
galaxy's interstellar medium as derived from various observational studies.
I\,Zw\,18 is a low mass galaxy which is embedded in a large \HI\ envelope 
\citep{van98}.  Within this neutral gas, we see in deep optical and infrared 
imaging (HT95\nocite{hun95}; DeMello et al. 1998\nocite{dem98}; Aloisi et al. 
1999\nocite{alo99}; \"Ostlin 2000\nocite{ost00}) the evolutionary signatures of 
stars down to the lower mass limit which is currently detectable ($\sim$ 1.1 
\msun\ in NICMOS imaging; \"Ostlin 2000\nocite{ost00}).  The large ionized 
regions and intense \halpha\ emission evidence the UV flux of massive stars and 
suggest ongoing star formation.  Mixed throughout the galaxy we find large dust 
concentrations, which could be due to either (or both) mass loss from massive 
stars or intermediate mass (AGB) stars that have been recently supported by the 
aforementioned CMD studies.   Regardless of its origin, however, the presence of
dust suggests that areas rich in molecular gas are abundant even at this very 
low metallicity.  The FUSE non-detection of diffuse H$_2$ in absorption 
\citep{vid00} may still be consistent with the presence of large amounts of 
molecular gas in I\,Zw\,18, since in these observations we find bulk dust 
distributions displaced from UV clusters and not along lines of sight directly 
to them.  We favor the scenario that the molecular clouds in I\,Zw\,18 (and 
likely in other low-metallicity galaxies) are small and dense.  This will cause 
them to be very difficult to detect, yet still allow them to actively and 
efficiently form the short-lived massive stars which we see throughout 
I\,Zw\,18.  

\acknowledgments 
The authors appreciate the thorough and insightful comments of an anonymous 
referee which helped to improve the manuscript.  Support for this work was 
provided by NASA through grant number GO-6536 from the Space Telescope Science 
Institute, which is operated by AURA, Inc., under NASA contract NAS5-26555, and 
by LTSARP grant NAG5-9221.  J.\,M.\,C. is supported by NASA Graduate Student 
Researchers Program (GSRP) Fellowship NGT 5-50346.  D.\,R.\,G. acknowledges 
support from NASA LTSARP grant NAG5-7734. This research has made use of the 
NASA/IPAC Extragalactic Database (NED) which is operated by the Jet Propulsion 
Laboratory, California Institute of Technology, under contract with the National
Aeronautics and Space Administration.

\clearpage

\clearpage

\clearpage

\clearpage
\begin{deluxetable}{cccrccl}
\tabletypesize{\scriptsize}
\tablecaption{HST WFPC2 Observations of I\,Zw\,18}
\tablewidth{0pt}
\tablehead{\colhead{Filter} & \colhead{Emission} & \colhead{Proposal}   
& \colhead{Date}   & \colhead{Exposure} &\colhead{WFPC2}  & \colhead{Dataset}\\
&\colhead{} &\colhead{} &\colhead{} &\colhead{(sec)} &\colhead{Chip} &\colhead{}}
\startdata
F487N 		&\hbeta		&GO-6536	
& 2/16/99 &2500	&WF3	&U39A0101R,U39A0102R,U39A0103R\\
F658N 		&\halpha	&GO-6536			
& 4/04/98 &2500	&PC	&U39A0201R,U39A0202R,U39A0203R\\
F658N 		&\halpha	&GO-5434			
&11/02/94 &4600	&WF3	&U2F90205T,U2F90206T\\
F450W 		&Wide-B		&GO-5434			
&11/03/94 &4600 &WF3	&U2F90102T,U2F90103T\\
F675W 		&R  		&GO-5434			
& 3/01/95 &2000 &PC	&U2F90305T,U2F90306T\\
F702W 		&Wide-R		&GO-5434			
&11/03/94 &5400 &WF3	&U2F90101T,U2F90201T,U2F90202T\\
\enddata
\label{t1}
\end{deluxetable}

\clearpage
\begin{deluxetable}{lcccccccc}
\tabletypesize{\scriptsize}
\tablecaption{Narrow Band Photometric Calibration Values}
\tablewidth{0pt}
\tablehead{\colhead{Filter} &\colhead{Emission} &\colhead{WFPC2}     
&\colhead{PHOTFLAM\tablenotemark{a}} &\colhead{URESP\tablenotemark{a}} 
&\colhead{Filter} &\colhead{Efficiency}	&\colhead{Continuum}	
&\colhead{Scaled\tablenotemark{b}} \\ &\colhead{} &\colhead{Chip} 
&\colhead{} &\colhead{} &\colhead{Width (\AA)}
&\colhead{Correction}	&\colhead{Filter}	&\colhead{Fraction}} 
\startdata
F487N  &\hbeta 		&WF3	&3.89$\times$10$^{-16}$    
&3.94$\times$10$^{-16}$    &33.9	&1.07	&F450W		&2.7\%\\
F658N  &\halpha 	&PC 	&1.06$\times$10$^{-16}$    
&1.04$\times$10$^{-16}$    &39.2	&1.01	&F702W		&1.9\%\\
F658N  &\halpha		&WF3	&1.06$\times$10$^{-16}$    
&1.04$\times$10$^{-16}$    &39.2	&1.01	&F675W		&2.9\%\\
\enddata
\tablenotetext{a}{Units of erg sec$^{-1}$ cm$^{-2}$ {\AA}$^{-1}$}
\tablenotetext{b}{Fraction which narrow band image was multiplied by before 
subtraction from wide band filter.  This value was found both iteratively and by
finding the ratio of (filter throughput at the wavelength of redshifted 
emission) to (filter width at full width half maximum); both ratios were 
calculated using the SYNPHOT package.  Detailed discussion of the continuum 
subtraction process is given in \S~\ref{S2.2}.}
\label{t2}
\end{deluxetable}

\clearpage
\begin{deluxetable}{lcccccc}
\tabletypesize{\scriptsize}
\tablecaption{The Bright \HII\ Regions of I\,Zw\,18}
\tablewidth{0pt}
\tablehead{\colhead{Feature}        
&\colhead{Center RA}     
&\colhead{Center DEC}  
&\colhead{Previous}            
&\colhead{Radius} 
&\colhead{L$_{\rm H\alpha}$}
&\colhead{O8 Ionizing}\\&\colhead{(J2000)} 
&\colhead{(J2000)} 
&\colhead{Identification\tablenotemark{a}}
&\colhead{(pc)} 
&\colhead{(10$^{36}$ erg sec$^{-1}$)}
&\colhead{Equivalent\tablenotemark{b}}}
\startdata
NW 1 	&9:34:02.361	&55:14:26.27	&HT 2	 &11.2	&511 	&238\\
SE 1	&9:34:02.610	&55:14:21.45	&HT 16	 &7.5	&112.8	&53\\
NW 2 	&9:34:02.428	&55:14:26.11	&HT 3	 &6.0	&71.1 	&33\\
SE 2	&9:34:02.680	&55:14:21.75	&HT 5	 &5.0	&42.3	&21\\
SE 3	&9:34:02.660	&55:14:20.35	&HT 13	 &3.4	&37.7 	&18\\
SE 4	&9:34:02.473	&55:14:20.73	&HT 10	 &4.5	&35.0 	&16\\
SE 5	&9:34:02.549	&55:14:22.39	&HT 6	 &2.5	&29.7 	&14\\
SE 6	&9:34:02.680	&55:14:20.62	&HT 14	 &3.0	&29.1 	&14\\
SE 7	&9:34:02.526	&55:14:22.13	&HT 7	 &3.00	&23.0 	&11\\
SE 8	&9:34:02.604	&55:14:20.84	&HT 12	 &3.0	&21.7 	&10\\
SE 9	&9:34:02.565	&55:14:20.92	&HT 11	 &2.9	&18.1 	&8\\
SE 10	&9:34:02.452	&55:14:21.34	&HT 9	 &2.4	&17.1 	&8\\
SE 11	&9:34:02.416	&55:14:21.49	&HT 8	 &2.6	&16.2 	&8\\
SE 12	&9:34:02.525	&55:14:20.98	&HT 11	 &2.5	&12.5 	&6\\
SE 13	&9:34:02.507	&55:14:21.78   &\nodata &2.85	 &9.2 	&4\\
Total Galaxy &9:34:02.0 &55:14:24.0 &\nodata	 &\nodata &6200 &2886\\
\enddata
\tablenotetext{a}{HT = \citet{hun95} = HT 95\nocite{hun95}}
\tablenotetext{b}{Assumes an ionizing photon rate of 10$^{48.6}$ sec$^{-1}$ 
for an O8 star \citep{ost89}.}
\label{t3}
\end{deluxetable}

\clearpage
\begin{deluxetable}{lcccccc}
\tabletypesize{\scriptsize}
\tablecaption{Polygonal Aperture Flux Ratios and Extinction}
\tablewidth{0pt}
\tablehead{\colhead{Feature}      &\colhead{\halpha\ Flux}              
&\colhead{\hbeta\ Flux}            &\colhead{\hab}	
&\colhead{A$_{\rm V}$\tablenotemark{a}}
&\colhead{\halpha\ Equivalent}		&\colhead{\hbeta\ Equivalent}\\
&\colhead{(erg sec$^{-1}$ cm$^{-2}$)} &\colhead{(erg sec$^{-1}$ cm$^{-2}$)} 
&\colhead{Ratio}			  &\colhead{(Magnitudes)}
&\colhead{Width\tablenotemark{b} \hspace{0.2 cm}(\AA)}	
&\colhead{Width\tablenotemark{b} \hspace{0.2 cm}(\AA)}}
\startdata
NW D1           &3.88$\times$10$^{-14}$ &1.34$\times$10$^{-14}$  
&2.88$\pm$0.02	&0.05$\pm$0.01	&908$\pm$3	&88$\pm$1\\
NW D2           &7.13$\times$10$^{-15}$ &2.49$\times$10$^{-15}$  
&2.87$\pm$0.04	&0.04$\pm$0.02	&2055$\pm$20	&192$\pm$3\\
NW D3           &2.05$\times$10$^{-14}$ &6.45$\times$10$^{-15}$  
&3.17$\pm$0.03	&0.29$\pm$0.02	&903$\pm$4	&75$\pm$1\\
NW D4           &3.02$\times$10$^{-14}$ &1.07$\times$10$^{-14}$  
&2.83$\pm$0.02	&0.00$\pm$0.01	&1527$\pm$7	&156$\pm$1\\
NW D5           &4.87$\times$10$^{-14}$ &1.65$\times$10$^{-14}$  
&2.96$\pm$0.02	&0.12$\pm$0.01	&1342$\pm$4	&140$\pm$1\\
NW D6           &4.17$\times$10$^{-15}$ &1.37$\times$10$^{-15}$  
&3.04$\pm$0.06	&0.19$\pm$0.03	&1412$\pm$15	&137$\pm$3\\
SE D1           &9.59$\times$10$^{-15}$ &3.20$\times$10$^{-15}$  
&3.00$\pm$0.04	&0.15$\pm$0.02	&920$\pm$6	&95$\pm$1\\
SE D2           &5.40$\times$10$^{-15}$ &1.93$\times$10$^{-15}$  
&2.80$\pm$0.04	&0.00$\pm$0.02	&1067$\pm$ 9	&126$\pm$2\\
SE D3           &6.81$\times$10$^{-15}$ &2.48$\times$10$^{-15}$  
&2.75$\pm$0.04	&0.00$\pm$0.03	&1175$\pm$12	&150$\pm$3\\
SE D4           &1.82$\times$10$^{-15}$ &5.35$\times$10$^{-16}$  
&3.40$\pm$0.11	&0.47$\pm$0.05	&1255$\pm$20	&101$\pm$3\\
SE D5           &9.66$\times$10$^{-15}$ &3.07$\times$10$^{-15}$  
&3.15$\pm$0.04	&0.28$\pm$0.02	&1369$\pm$10	&115$\pm$2\\
SE D6           &1.12$\times$10$^{-14}$ &3.73$\times$10$^{-15}$  
&3.00$\pm$0.04	&0.15$\pm$0.02	&1110$\pm$7	&101$\pm$1\\
SE D7           &5.51$\times$10$^{-15}$ &1.87$\times$10$^{-15}$  
&2.94$\pm$0.05	&0.10$\pm$0.03	&1103$\pm$9	&103$\pm$2\\
SE D8           &4.72$\times$10$^{-15}$ &1.56$\times$10$^{-15}$  
&3.03$\pm$0.06	&0.18$\pm$0.03	&767$\pm$6	&84$\pm$2\\
\enddata
\label{t4}
\tablenotetext{a}{Corrected for 0.06 magnitudes of foreground extinction.}
\tablenotetext{b}{Average value over the aperture, rounded to nearest \AA.}
\end{deluxetable}
\end{document}